\documentclass[thmsa,10pt]{article}
\begin{document}

\title{Exploring effective interactions 
through transition charge density study of $^{70,72,74,76}\rm Ge$ nuclei.}

\author{A. Shukla, P. K. Raina, P. K. Rath $^{a}$ \\
Department of Physics and Meteorology, IIT Kharagpur-721302, India.\\
$^{a}$Department of Physics, University of Lucknow, Lucknow-226007, India.\\}

\maketitle

\begin{abstract}

Transition charge density (TCD) for $\ 0^+\to 2_1^+ $ excitation have been
calculated for $\rm ^{70,72,74,76}Ge$ nuclei within microscopic variational
framework employing $2p_{3/2},~ 1f_{5/2},~ 2p_{1/2}$ and $1g_{9/2}$ valence
space. The calculated TCDs for different monopole variants of Kuo interaction
are compared with available experimental results. Other systematics like
reduced transition probabilities B(E2) and static quadrupole  moments Q(2) are
also presented. It is observed that the transition 
density study acts as a sensitive probe for discriminating the response of different parts 
of effective interactions. \\

Keywords : Hartree Fock Bogoliubov, Effective Interactions, Transition charge density \\

Pacs Nos: 21.60.-n, 21.10.-k, 23.20.-g, 27.50.+e \\

\end{abstract}

\section{INTRODUCTION}

Inelastic electron scattering experiments restricted by low beam energies, 
provided much useful information on spins of particles, moment of the nuclear
transitions such as electromagnetic transition probabilities and transition
radii. Most of   the electron scattering results were interpreted in terms of
macroscopic models with a common feature of transition densities having
surface peak. There is no a priori reason for the transition
density to have a surface peaked shape, rather, the   existence of interior
peak may be a stringent test for any microscopic calculation. The   advent of
high-resolution techniques has made it possible to observe the transition 
densities for different excitations in the interior of nucleus \cite{[heisen]}.
It has opened up a new and powerful way to check the nuclear structure
calculations with high precision. Recently, Richter 
\cite{rich00} and Neumann-Cosel \cite{Cos01} have provided
a state-of-the-art of inelastic electron scattering experiments and its potential to
explore the subtle nuclear structure effects having close relations to the 
key problems of nuclear physics. Very recently Radhi \cite{radh02} has also shown 
that this technique helps much in discriminating the finer effects of two body interactions
and core-polarisation even in s-d shell nuclei when shape transitions become 
important. 

Shell model has a potential to explore the physical phenomenon
behind some unexpected experimental observations in different nuclear properties.
One can have microscopic understanding of the dynamics through comparison of
experimental results with the model calculations having specific
varied  ingredients. An example where such a study has been carried through
is that of TCD for even-even Ni nuclei. Yokoyama and Ogawa \cite{[yoko89],[yoko90]}
carried out a shell model study within in ($2p_{3/2}, 1f_{5/2}, 2p_{1/2})^n$
configuration. Working with three different effective interactions and looking
at   the effect of p-h excitation from ($1f_{7/2})$ they concluded that the
effective charge model   which gave satisfactory description of BE(2) and Q(2)
values, completely failed in   describing the TCD. One gets an interior peak
that is some times even larger than   surface peak. They could analyse the
microscopic origin of peak by looking at the   contribution of $0 \hbar 
\omega$ and $2 \hbar \omega$ separately. The phases of individual
contributions to TCD   in these cases for the interior peak were found to be
opposite than that for the surface   peak. Thus the combined result leads to
desired experimental behaviour of small peak inside and large peak at the
surface. This phase dependence of the TCD contributions from different
intrinsic phenomenon makes their study sensitive. 

Two different approaches have been evolved in literature to handle nucleon-nucleon 
(N-N) effective interactions within  many body interaction framework for 
microscopic studies of nuclear properties.  One of them, initiated more than thirty
years ago by Kuo and Brown \cite{kuo68} is based on the  different N-N forces and the 
regularization methods. This technique has been
developed and applied in computing the effective 
 interactions in different
shells and is still being pursued actively. On the other hand, the possibility of 
separation of nucleonic 
interaction Hamiltonian into multipole fields too has been of quite interest
from earliest 
 times \cite{kiss63,bara68,kish72,bohr75}. Some recent studies 
\cite{hase01,hase00} have attempted to identify the importance of lower
multipoles and their strengths through P+QQ interactions and its extensions.
They have been quite successful. 

It has been noticed \cite{abzo91} that the nuclear Hamiltonian can be separated 
rigorously in to monopole and multipole fields. Monopole fields with few parameters 
describe saturation properties and spectroscopic properties very well. Apart from their success
\cite{pove91,caur99,koon97} in describing binding energies and 
 other global
properties like M1, E2 and GT sum rules, they are also being used \cite{caur99}
as benchmarks for new promising approximations of solving large scale shell
model problems in Monte Carlo methods \cite{koon97,honma96}. Monopole modifications have also
been  applied in f-p-g shell for the studies of double beta decay matrix
elements of $\ ^{76}\rm Ge$ and   $\ ^{82}\rm Se$ nuclei in variational
framework \cite{dhiman94} and shell model \cite{caur96} apart from electron capture
rates in astrophysics.

Hasegawa, Kaneko, Tazaki and Zhang have shown \cite{hase01,hase00} recently
that P+QQ 
 empirical effective interactions are very close to the realistic
interactions provided J-independent proton-neutron (p-n) force is included
properly. These studies have shown 
 this equivalence explicitly in case of
some p-f shell nuclei that is the best place \cite{hase00} for 
 such comparison.
In case of medium mass (mass region 100) nuclei where shell model 
 becomes
unmanageable even with recent computation facilities there have not been 
much of the tested full shell effective interactions. So multipole field
interactions have 
 been the natural choice for microscopic variational models
like Hartree-Fock, Hartree-Fock-Bogolioubov as well as bosonic models 
(different versions). There have been 
 studies \cite{bren90,singh94}, attempting to
evolve some general trends of p-n interactions, which 
 have led to the
important conclusion that T=0 component of the interaction plays 
 crucial
role in describing microscopic properties. In particular, transition charge
density 
 studies \cite{singh94} have acted as a sensitive way for fixing p-n
component of Q.Q interaction.

Here we wish to demonstrate how this technique is very useful in 
discriminating between the role of different components of interactions.
Inelastic electron scattering data in
Germanium nuclei has shown some very interesting features \cite{[bazant]} that have
been explained in configuration mixing framework \cite{[duv83]} of   interacting
boson model (IBM). There have been some microscopic attempts for the 
description of transition charge densities in various mass regions
\cite{[heisen],[bazant],[schwen],[wessel],[millim],[raina88],[yoko89],[yoko90]}.   Theoretical attempts in
IBM, Shell model and Hartree-Fock-Bogoliubov (HFB) model   have been made in
literature to quite some extent with IBM acting as a convenient   rather than
microscopic framework. We present the 
 theoretical
study of transition charge densities for quadrupole excitation ($0^+\to
2^+_1$) in 
 case of $\ ^{70,72,74,76}\rm Ge$ nuclei in microscopic variational
framework. Sensitivity of 
 the first peak to specific variation of monopole
strengths that result into the desired 
 experimental observations, helps to
identify the proper type and strengths of monopole 
 changes on effective Kuo
interactions in f-p-g valence space. This region which includes nuclei 
 like Zn, Ge,
Se, Kr, Sr and Zr show many interesting features from microscopic 
 structure
aspects in terms of shape transitions as well as deformations even in low
lying states. These nuclei have been studied within variational models, 
restricted 
shell models and
Bosonic models. We find that the study of transition 
 charge
densities is probably the finer way to explore effective interactions than
other 
 electromagnetic properties like spectra, BE(2) and Q(2) that have been
extensively used in the literature.

\section{CALCULATIONAL FRAMEWORK}

The formalism to calculate the wavefunction $\vert\Phi\rangle$\  is based
upon the Hartree-Fock-Bogoliubov  (HFB) Method. The HFB theory generalizes and
unifies the Hartree-Fock (HF) procedure (with HF field) and the BCS Model (with
pairing field) by treating them simultaneously on equal footing. 
The two-body hamiltonian H is given by
\begin{eqnarray} {H}=\sum\limits_{\alpha}\epsilon_{\alpha} a_{\alpha}^\dagger
a_{\alpha} + \frac{1 }{ 4}\sum\limits_{\alpha\beta\gamma\delta}
\langle\alpha\beta|V|\gamma\delta\rangle a_{\alpha}^\dagger a_{\beta}^\dagger
a_{\delta} a_{\gamma}\end{eqnarray}
Axially symmetric HFB intrinsic
state with K=0 can be written as
\begin{eqnarray} {|\Phi_0\rangle}=\prod\limits_{im}(U_{im} +
V_{im}b_{im}^\dagger b_{i\bar m}^\dagger)|0\rangle\end{eqnarray}
where the creation operators $\ b_{im}^\dagger$\ and $\ b_{i\bar
m}^\dagger$\ are given by 
\begin{eqnarray} {b_{im}^\dagger}=\sum\limits_{\alpha}
c_{i\alpha ,m}a^{\dagger}_{\alpha m}\hskip0.5in {\rm and}\hskip0.5in b_{i\bar
m}^\dagger=\sum\limits_{\alpha} (-1)^{j-m}c_{i\alpha ,m}a^{\dagger}_{\alpha,
-m}\end{eqnarray} Using the standard projection technique, a state with
good angular  momentum   is obtained from the HFB intrinsic state through the
relation.

\begin{eqnarray} {|\Psi_{MK}^J\rangle}&&=P_{MK}^J|\Phi_K\rangle\nonumber\\
&&=\left[\frac{(2J+1) }{ {8\pi^2}}\right]\int
D_{MK}^J(\Omega)R(\Omega)|\Phi_K\rangle d\Omega\end{eqnarray}
where $\ R(\Omega)$\ and $\ D_{MK}^J(\Omega)$\ are the rotation operator and
the rotation matrix respectively. 

\subsection {Static electromagnetic properties }Expressions used to calculate 
reduced transition probabilities BE(2) and static quadrupole moments $ Q(J^\pi)$  are
given below.

Employing the angular momentum projected wave-function  $|\Psi_{K}^J\rangle$, 
one obtains the following  expression for reduced transition probability B(E2)
\begin{eqnarray}{B(E2:J_i\to J_f)}=\left( \frac{5 }{16\pi}\right)\left(
e_\pi\langle Q_{0}^2\rangle _{\pi} + e_\nu\langle Q_{0}^2\rangle
_{\nu}\right)^2\end{eqnarray}
where 
\begin{eqnarray}{\langle
Q_{0}^2\rangle _{\tau_3}}&&=\langle\Psi_{K}^{J_i}\
|Q_{0}^2|\Psi_{K}^{J_f}\rangle\nonumber\\
&&=\left[ n^{J_i}n^{J_f}\right]^{-1/2}\int\limits_{0}^\pi\sum\limits_{\mu}
\left(\begin{array}{ccc}J_i&2&J_f\\-\mu&\mu&0\end{array}\right)d_{-\mu0}^{J_i}(\theta) n(\theta)
\nonumber\\ && \times\left[
b^2\sum_{\tau_3\alpha\beta}e_{\tau_3}\langle\alpha|Q_{\mu}^2|\beta\rangle
\rho_{\alpha\beta}^{\tau_3}(\theta)\right]sin\theta\ d\theta
\nonumber\\&&\end{eqnarray} 
\begin{eqnarray}{n(\theta)}=\sqrt{det[1+M(\theta)]}\end{eqnarray} \begin{eqnarray}
{M(\theta)}=F_{\alpha\beta}(\theta)f_{\alpha\beta}^{\dagger}\end{eqnarray}
\begin{eqnarray}
\label{eq23}F_{\alpha\beta}(\theta)=\sum\limits_{m'_{\alpha}m'_{\beta}}
d^{j_{\alpha}}_{m_{\alpha},m'_{\alpha}}(\theta)d^{j_{\beta}}_{m_{\beta},m'_{\beta}}(\theta) 
f_{j_{\alpha}m'_{\alpha},j_{\beta}m'_{\beta}}\end{eqnarray} 
\begin{eqnarray}\label{eq24}
{f_{\alpha\beta}}=\sum\limits_{i}c_{ij_{\alpha},m_{\alpha}}
c_{ij_{\beta},m_{\beta}}\left(V_{im_{\alpha}}/U_{im_{\alpha}}\right)
\delta_{m_{\alpha} ,-m_{\beta}}\end{eqnarray}\begin{eqnarray}{n^{J}}=\int\limits_{0}^{\pi}
n(\theta) d_{00}^J(\theta)sin\theta d\theta\end{eqnarray}
\begin{eqnarray}
{\rho_{\alpha\beta}^{\tau_3}(\theta)}=\left[M(\theta)/(1+M(\theta))\right]
_{\alpha\beta}^{\tau_3}\end{eqnarray}  and\begin{eqnarray} {Q_{\mu}^2}={\sqrt
\frac{16\pi }{ 5}} \frac{r^2 }{ b^2} Y_{\mu}^2(\theta,\phi)\end{eqnarray}

Similarly the static quadrupole moments $\ Q(J^\pi)$\ are evaluated using the 
expression.
\begin{eqnarray} {Q(J^\pi)}&&=\langle\Psi_{K}^J|Q_{0}^2|\Psi_{K}^J\rangle\nonumber\\
&&=\left[ n^{J}\right]
^{-1}\left(\begin{array}{ccc}J&2&J\\
J&0&J\end{array}\right)\int\limits_{0}^\pi\sum\limits_{\mu}
\left(\begin{array}{ccc}J&2&J\\
-\mu&\mu&0\end{array}\right)\nonumber\\&&\times \hskip1mm d_{-\mu0}^J(\theta)
n(\theta) \left[
b^2\sum_{\tau_3\alpha\beta}e_{\tau_3}\langle\alpha|Q_{\mu}^2|\beta\rangle
\rho_{\alpha\beta}^{\tau_3}(\theta)\right]sin\theta\ d\theta\end{eqnarray} 

\subsection{Transition Charge Density}

The TCD, $\rho_{L}(r)$\ is the reduced matrix element of $\rho_{L}^{op}$\ 
between the initial and  the final nuclear state of spin $\ J_i$\ and $\ J_f$\ 
and is given by \begin{eqnarray}
\rho_{L}(r)=\langle\Psi_{K}^{J_f}\vert\vert\rho_{L}^{op}\vert\vert
\Psi_{K}^{J_i}\rangle\end{eqnarray}  
Employing the HFB wave functions, one obtains the following expression for TCD
\begin{eqnarray}\langle\Psi_{K}^{J'}\vert\vert
\rho_L^{op}\vert\vert\Psi_{K}^J\rangle && = \left[n^Jn^{J'}\right]^{-\frac{1}{2}}\frac
{(2J+1)}{2}\int_{0}^{\pi/2}\sum_\mu\left(\begin{array}{ccc}J&2&J'\\-\mu&\mu
&0\end{array}\right)d_{-\mu0}^J(\theta)\nonumber \\ && \times n(\theta)\left[
b^2\sum_{\tau_3\alpha, \beta\ }e_{\tau_3}R_{n_{\alpha}l_{\alpha}}(r)
R_{n_{\beta}l_{\beta}}(r)\langle\alpha\vert Y_M^L \vert\beta\rangle
(\theta)\right] \rho_{\alpha\beta}^{\tau_3}sin\theta d\theta \end{eqnarray}
with \begin{eqnarray}\rho_{\alpha\beta}^{\tau_3}=
\left(M(\theta)\left[1+M(\theta)\right]^{-1}\right)_{\alpha\beta}^{\tau_3}
\end{eqnarray} and \begin{eqnarray} {n^{J}}=\int\limits_{0}^\pi \left[det\left(
1+F^{(\pi)}f^{(\pi)^{\dagger}}\right)\right]^{1/2}\left[det\left(
1+F^{(\nu)}f^{(\nu)^{\dagger}}\right)\right]^{1/2}
d_{00}^J(\theta)sin(\theta)d\theta\end{eqnarray} 
$\it b$ is the oscillator length parameter,
$R_{n_{\alpha}l_{\alpha}}(r)$ are harmonic oscillator  wave-functions and  $e_{\tau _3}$ is
effective charge.

The calculations have been performed with Kuo effective interaction operating in
the valence space spanned by $\ 2p_{3/2}$, $\ 1f_{5/2}$, $\ 2p_{1/2}$ and
$\ 1g_{9/2}$ orbits. The doubly closed nucleus $\rm ^{56}Ni$\ is treated as an
inert core. The relevant effective two-body interaction that we have employed
is a renormalized G-matrix due to Kuo \cite{kuo}. The single particle energies taken
(in MeV) are $\epsilon(2p_{3/2})=0.00$, $\epsilon(1f_{5/2})=0.78$,
$\epsilon(2p_{1/2})=1.08$ and $\epsilon(1g_{9/2})=3.50.$\ 

These interactions have been used for satisfactory explanation of the 
observed anomalous high-spin sequence in $^{60}\rm Ni$ with shell model
calculations as well as for the theoretical studies of electromagnetic
properties of the yrast and yrare states in Zn, Ge, Se and Kr isotopes \cite{ahalp82}.
Quite extensive studies of static properties of 
 Germanium and Selenium
isotopes in the microscopic variational framework have been 
 reported \cite{trip86,rath88}.
We present the results of projected 
 HFB calculations for the transition
charge densities, transition probabilities BE(2) and static quadrupole moment Q(2) 
for even-even $^{72-76}\rm Ge$ 
 nuclei in next
section and discuss how transition charge density calculations can be 
 used in
identifying monopole strengths.

\section{RESULTS AND DISCUSSION}
Two-body interaction matrix elements used in the calculation of 
electromagnetic and weak interaction properties, as discussed in the introduction, form 
a very essential and most important input to any nuclear model calculations. Let us 
look at the evolution of these two body interactions in the
f-p valence space.  From the literature we can notice that the use of two
body interactions has been of  considerable interest to many theoretical
attempts in f-p and f-p-g shell for the  description of energy spectrum,
multipole moments and the transition probabilities for  more than three
decades. The realistic interaction for the f-p shell nuclei was 
constructed by Kuo and Brown \cite{kuo68} and put to test successfully for extracting the spectra of some nuclei near sh
ell clsoure. Then
exhaustive spectroscopic shell model 
 calculations for Ca nuclei by McGrory,
Wildenthal and Halbert \cite{mcgr70} revealed that these 
 interactions have to
be modified slightly suggesting that the effective interaction for $\ f_{7/2}$
with other orbits of the space are too strong. The next attempt made in
modifying 
 these effective interactions in f-p region was made by Sharma and
Bhatt \cite{sharma73} to 
 examine the intrinsic structure of even-even nuclei of
Ti, Cr, and Fe.  Based on these 
 results and Nilsson structure of the orbits
they suggested, in line with the MWH, that 
 matrix elements for particles
in $(\ f_{7/2})^2$ should be made more attractive and the 
 interaction of $\
f_{7/2}$ with other orbits should be made more repulsive. The variation of 
these interaction matrix elements has been tried by them for 100, 200 and 300
KeV.  
 Next very effective step in this development was that of monopole
modifications \cite{pove91}. 
 The tremendous success of this modification in
description of global properties and now using them for benchmarking of
promising new techniques (like Monte Carlo) has established it as KB3  interaction. 
  With the success of 
 monopole
modifications in f-p region, there have been now universal acceptability for 
them. Similar monopole modifications have been recently applied to the studies
of 
 double beta decay matrix elements of $\ ^{76}\rm Ge$ and  $\ ^{82}\rm Se$
nuclei in variational framework 
 \cite{dhiman94} and shell model \cite{caur96} employing 
f-p-g valence space.
These are still considered to be one of the best microscopic 
 nuclear matrix
elements calculations in literature with the shell model nuclear matrix 
 elements
for neutrinoless double beta decay being taken as most reliable ones till
date. 

Ge nuclei have long presented challenge for microscopic description because
of 
 anomalous behaviour in energy systematics as well as transition charge
densities of 
 ground/low-lying states. There have been some discontinuous 
changes between
neutron number 
 N=40 to 42. IBM plus configuration mixing \cite{[duv83]} has been
applied successfully to the 
 studies of systematics as well as transition
charge densities. In these studies the idea of 
 two different configuration
has been evolved because of crucial role played by $\ p_{3/2}$ and 
 the $\
f_{5/2}$ orbits. The calculations with restricted shell model 
\cite{xi89,list90,hend97}, Variational
models \cite{ahalp82,trip86}  and IBM \cite{[bazant],[duv83]} carried out in f-p-g region
have shown that $\ f_{5/2}$ orbits are playing  very crucial role in
describing static and dynamic properties. Truncated shell model calculations for N=50
(Zn-Rb) isotones 
 in   f-p-g valence space too indicate \cite{xi89} that the
occupancies of orbits are mainly 
 dominated by $\ f_{5/2}$ orbit followed by
$\ p_{3/2}$. For lower mass region covering Zn to Zr 
 nuclei it is found that
the mixing with $\ 1f_{5/2}$ and $\ 2p_{3/2}$ orbits is very strong
\cite{list90} and 
 truncations excluding these levels may be justified only
from mass 86 onwards.  Very recently Langanke,
Kolbe and Dean proposed \cite{lang01} a new model to calculate stellar electron
capture rates using the Shell Model Monte Carlo approach for even mass germanium
isotopes in f-p-g valence space with pairing+quadrupole interaction. 
They adopted the single-particle energies from the KB3 interaction but had to
artificially reduce the $\ f_{5/2}$\ orbit by 1 MeV to simulate the
effects of the $\sigma \tau$ component that was missing in the residual interaction. 
This too points towards the need for larger occupancy of $\ f_{5/2}$\ orbit.

Taking a clue about the role of important orbits from these studies and looking at the 
success of monopole modifications in f-p region, we have attempted  (restricting to 
yrast state excitations) transition charge densities for all Ge nuclei that
have been studied experimentally.
Though yrare  excitations too are very interesting  
 but are believed to be
connected to the shape transitions and here we have confined 
 ourselves with the
studies of monopole effects. Specifically, to the identification of important
orbits and finding their appropriate strengths. We explore the effect
of monopole modification of effective two-body 
 interaction with variation of
strength by 100, 150, 200 and 250 keV.  Kuo \cite{kuo} interaction 
 is
denoted by Kuo00 effective interaction. Kuo00 has been modified by making
$\langle (f_{5/2})^2 
 JT| V |(f_{5/2})^2 JT\rangle$ interaction matrix
elements attractive by  100 keV and $\langle (p_{3/2})^2 JT| V 
 |(p_{3/2})^2
JT\rangle$  interaction matrix elements repulsive by same amount (this
modification is 
 called Kuo10). Replacing 100 by 150, 200 and 250, we get
Kuo15, Kuo20 and Kuo25 
 respectively. Detailed studies of transition
charge density for first quadrupole excitation along with the static 
electromagnetic properties with
respect to the variation   of these monopole strengths in case of even-even
$^{70-76}\rm Ge$ nuclei  are presented in following subsections. We have
tried different monopole modifications with different orbits as well as
same/different strength variations for all orbits of the space, including the
effect of transitions from/to $p_{1/2}$ as well as $g_{9/2}$ orbits. It is
found that the combination that has been chosen is the best one to reproduce the
desired characteristic of both peaks of TCDs simultaneously.
\subsection{Static Electromagnetic Properties}
\subsection*{\it 1. Total energy, Intrinsic Quadrupole Moment and Occupation
Numbers}
 
Table I presents the total HFB energy and total quadrupole moments
(along with 
 separate contribution from protons and neutrons) for
$ ^{70,72,74,76}\rm Ge$ nuclei. Also presented 
 are their variations with
changes in monopole strengths discussed above. 
$^{70,72}\rm Ge$ nuclei show first increase in quadrupole moment and then
decrease 
 as we go from Kuo00 to Kuo25. Whereas, $ ^{74,76}\rm Ge$ nuclei
show decreasing trend 
 except for $^{74}\rm Ge$ at Kuo00 to Kuo10 where it
hardly shows any change.
Table II shows the change in proton and neutron occupation numbers in 
$ 2p_{3/2}, ~1f_{5/2},~ 2p_{1/2}$ and $1g_{9/2}$ orbits as we go on increasing
monopole strength from 000 to 
 250 in steps. As expected we find
promotion of particles from $p_{3/2}$  to $f_{5/2}$ orbits 
 both for protons
and neutrons.  
We notice that the occupancy of protons is almost completely governed by
$1f_{5/2}$ orbit 
 with little bit spreading into $2p_{3/2}$ and $1p_{1/2}$ in
case of Kuo25. Whereas, at Kuo20 
 some of them shift to $2p_{3/2}$ and also
occupy little bit of $1g_{9/2}$ (seen in case of $\rm ^{70,72}Ge).$ 
Additional neutrons mainly keep occupying $1g_{9/2}$ in going from $^{70}$Ge
to $\rm ^{76}Ge.$ There is 
 not much change in neutron occupancies of $2p_{1/2}$  and
$1g_{9/2}$ orbits while going from Kuo00 to 
 Kuo25 and expected variation in
other two orbits is similar to those of protons. The 
 effects of these
transitions of neutrons and protons into different orbits have been 
reflected on TCDs, B(E2) and Q(2) discussed 
 in details in the following sections.

\subsection*{\it 2. Reduced Transition Probabilities B(E2) and Quadrupole Moments Q(2)}
The calculated as well as experimentally observed values for the reduced transition 
probabilities B(E2; $0^+\to 2_1^+$) and static quadrupole moments Q($2^+$) 
for $\rm ^{70,72}Ge$ are presented in Table I. The calculated results with 
effective charges $e_{eff}=e_\nu $=0.1 and 0.2 along with those for different variants
of Kuo00 are given in two columns. A closer look at the values shows that the changes 
from Kuo00 to Kuo25 in case of $^{70}$Ge increases BE(2) values in going from 
Kuo00 to Kuo10 and then keeps decreasing. In case of $^{72}$Ge nucleus the values at 
Kuo00 and Kuo10 are nearly same and then go on decreasing monotonically. The set of 
$^{74,76}$Ge nuclei have quite different behaviour from that of  $\rm ^{70,72}Ge$ in 
two respects. First, the values keep on decreasing very fast in going from Kuo00 to Kuo25. 
Secondly, there is a need for change in effective charge by approximately 0.3 units
to have the BE(2) values closer to those suggested by experiments. We find that the 
calculated values 
match well with experiments in case of Ku20/Kuo25 for all Ge nuclei (within the presented small 
variation of effective charges). Thus Kuo20/Kuo25 seems to be an appropriate monopole 
modification. We notice that the B(E2) values change very fast with e$_{eff}$ and thus one 
could also arrive at the agreement with experimental values without going for monopole 
modifications but changing effective charge for each nucleus. This is a normal practice 
followed where one treats e$_{eff}$ as free parameter in calculation and fixes it by 
experimental data. We shall see in next section that this kind of parameterization does 
not help with TCD studies. Theoretically calculated values for Q(2) along with experimental
results (with large error bars) are also tabulated in last three columns. Similar changes 
as seen for BE(2) are observed in this case too. Due to large error bars in experiments 
nothing conclusive can 
be said by comparison except that the qualitative trends are shown with the need for increment of e$_{eff}$ 
by 0.3 in going from $^{72}$Ge to $^{74}$Ge.

\subsection{Transition Charge Densities for $0^+ \to 2_1^+$
excitation} 
            
In figures 1(a) and 1(b) we have presented the
results for $0^+\to 2_1^+$ transition 
 charge densities of 
$^{70,72,74,76}\rm Ge$ nuclei. Experimental \cite{[bazant]} and the calculated 
results with unmodified Kuo (Kuo00) as well as for four modified
KB 
 interactions discussed above are also shown on different
curves of each
 figure. Experimental plots of transition charge densities
(between solid lines with 
 vertical bars standing for errors) for $0^+\to
2_1^+$  are characterized by one small peak in the 
 interior around r = 1.25
fm and a large surface peak at r = 4.25 fm. The surface peak in case of
$^{70,72}\rm Ge$ nuclei is smaller in comparison to $^{74,76}\rm Ge$
nuclei.

These experimental results have been attempted in IBM
framework which has 
 been quite successful in unified description of
collective states. To describe transition 
 densities in this framework one
has to introduce boson densities and most of the 
 experimental data goes as
input e.g. in reference 24 four sets of experimental data out 
 of total of
eight available sets were used as inputs. Whereas, shell model and variational
model  calculations have direct link with effective interactions; the changes
in occupancies of protons from a specific orbit to the different orbits are displayed
through transition charge density in a very sensitive way.

Figures 1(a) and 1(b) show the variation of transition charge densities for
$0^+\to2^+_1$  excitation in case of $^{70,72}\rm Ge$ and $^{74,76}\rm Ge$
nuclei respectively.  Here effective 
 charges for $^{70,72}\rm Ge$ and
$^{74,76}\rm Ge$ nuclei are taken to be $(e_\pi, e_\nu)$ =(1.15, 0.15) and
(1.45, 
 0.45) respectively so as to give the surface peak matching with
experimental values. In 
 case of iso-scalar $(e_\pi - e_\nu = 1.0)$ effective
charge interpretation one associates the effect 
 of non-zero effective charge $e_{eff}=e_\nu$
physically with core polarization. 
Thus the required change of 
effective charge by 0.3 in going from $^{72}\rm Ge$ to $^{74}\rm Ge$ shows the
need for more core  polarization in going from N=40 to 42 nuclei.  
This change is quite noticeable on experimental 
 results. Since N=40 is not a good core and 
the monopole modifications hardly affect surface peaks, it indicates some important
role being played by p-n interactions at N=40.

Now let us concentrate on the first peak. Transition charge densities with Kuo00,
Kuo10, Kuo15, Kuo20 and Kuo25 interaction are plotted with doted, dashed, crosses
on solid, 
circles on solid and triangles on solid curves respectively. Theoretical
results show that the 
 first peak is very large in case of Kuo00 interaction
(i.e. without any modification). 
 Very interesting feature of this peak is
its sensitivity to small variations in monopole 
 strengths of  orbits
$1f_{5/2}$ and $2p_{3/2}.$ As we go on increasing the monopole strength by
100keV and then in steps of 50 keV the peak goes on decreasing and
approaches towards experimental 
 behaviour very fast. This is a general trend
shown in case of all nuclei till Kuo20 
 though there appears to be somewhat
better correspondence with experimental results 
 even beyond 200 in case of
$^{70,72}\rm Ge$. As we go to $^{74}\rm Ge$ we find the peak almost 
 stagnated
at Kuo20, thus suggesting Kuo20/Kuo25
to 
be the appropriate monopole strength modification. Minima of
calculated values between two peaks is found to be 
 deviating from
experimental curve and other detailed overall differences are expected to be due to the 
finer effects of higher order contributions that have not been represented exactly in Kuo
interactions. In earlier elaborate
variational model calculations on   Ge nuclei \cite{tassie56,brown83} the computed BE(2) and Q(2)
values needed the variation of effective charges from nucleus to nucleus 
for better agreement with   experimental results. One important reason
for it appears to be the non-inclusion of appropriate   monopole strength.  

Even-even $^{64-68}\rm Zn$ nuclei studied \cite{[raina88]} earlier within full f-p
shell employing 
 MWH \cite{mcgr70} interactions displayed similar behaviour as
discussed above for Ge nuclei  with Kuo interaction. There it was understood as
a consequence of limiting valence 
 space that excluded $g_{9/2}$ orbit. Other
possible explanation for large interior peak was 
 believed to be the
contribution from higher multipole interactions. The purely empirical 
 Tassie
model \cite{tassie56, brown83}, where one adds core polarization transition charge
density 
 given by the derivative of ground state charge density, was seen to
provide better 
 fitting with the experiments. This purely empirical model
does not shed any light on  microscopic phenomenon responsible for suppression
of interior peak like the one 
 demonstrated \cite{[yoko90]} in case of TCD for Ni
isotopes. Our study 
 of TCD in case of
$^{64,66,68}\rm Zn$ nuclei, in same framework as discussed above for Ge 
nuclei, shows that these monopole modifications have a similar role to
play in 
 these nuclei too, thus showing that the contribution to quadrupole 
TCD amplitude due to
particles in $1p_{3/2}$ has same phase
for interior and the surface region
while it is in opposite phase in case of $1f_{5/2}$ orbit. Thus we can very clearly say
that the desired experimental behaviour for the TCDs of Ge nuclei in particular and
other neighbouring nuclei in general show dominant occupancy in $1f_{5/2}$ orbit.

\section{SUMMARY AND CONCLUSION}
 Microscopic variational model calculations
of transition charge density for 
 $0^+\to2^+$ excitation, transition probabilities 
B(E2) and static quadrupole moment Q(2) have been reported 
for $^{70,72,74,76}\rm Ge$ nuclei. The 
 valence
space $ 2p_{3/2}, ~1f_{5/2},~ 2p_{1/2}$ and $1g_{9/2}$ 
along with Kuo interaction has been used. The 
 successes of monopole
modifications on KB interactions (KB3 interaction) in f-p shell 
 has prompted
us to look for some better way to extract the strengths for such modifications 
 in
f-p-g shell. Germanium nuclei have long presented challenge for microscopic 
description because of anomalous behaviour in energy systematics as well as
transition  
 charge densities. IBM with two different configurations has been
successfully evolved to study 
 these anomalies where the role of mixing between 
$p_{3/2}$ and the $f_{5/2}$ orbits is found to be crucial. Shell model studies in this region too
have shown the importance of occupancies of $f_{5/2}$ orbits. 
 With these
observations in mind, we have examined the changes in transition charge 
densities with variation of strength by  100, 150, 200 and 250 keV for $\langle
(f_{5/2})^2   JT| V |(f_{5/2})^2 JT\rangle$ and $\langle (p_{3/2})^2   JT| V
|(p_{3/2})^2 JT\rangle$ interaction matrix elements. It is seen 
explicitly that the transition densities of inner peak are very sensitive to
these changes whereas the surface peak is almost inert. The desired variations 
in inner peak of transition   density so as to have 
observed experimental behaviour of small inner peak and large surface peak 
in case of $^{70,72,74,76}\rm Ge$ nuclei  suggest Kuo25/Kuo20 to be the
appropriate   monopole modification strength.


Table I. The calculated values of total
energy (in MeV), intrinsic quadrupole moment (in units of $\ b^2$), 
B(E2; $0^+\to 2^+_1)$ (in units of 
$e^2b^2$) and static quadrupole moment Q(2) (in units of $\it b$) for the
ground state HFB solution of $\ ^{70,72,74,76}\rm Ge$ nuclei. Here $\ \langle
Q^2_0\rangle_\pi$ and $\ \langle Q^2_0\rangle_\nu$ are separate contribution of
the protons and neutrons respectively. \label{tab1} 

\begin{tabular}{cccccccccccc}
\hline\hline

Nucleus&Interaction&$\ E_{HFB}$&$\ \langle Q^2_0\rangle_{HFB}$&$\ 
\langle Q^2_0\rangle_{\pi}$&$\ \langle Q^2_0\rangle_{\nu}$&
\multicolumn{3}{c}{B(E2; $0\to2_1^+$)}&\multicolumn{3}{c}{Q(2)} \\ \cline{7-9} \cline{10-12}
&&&&&&\multicolumn{2}{c}{Theory}&Expt.\footnotemark[1]&\multicolumn{2}{c}{Theory}&Expt.\footnotemark[2]\\  
&&&&&&$e_\pi$=0.1&0.2&&$e_\nu$=0.1&0.2&\\ 
\hline
& Kuo25&-27.29&28.49&10.8&17.7&18.41&22.76&17.90$\pm $0.30&-0.17&-0.19&-0.09$\pm $0.06\\
&Kuo20 &-26.47&29.69&11.4&18.3&19.71&24.46&17.70$\pm$4.60&-0.18&-0.20&\\
$\ ^{70}\rm Ge$&Kuo15 &-25.89&30.05&11.9&18.2&20.88&25.83&17.50$\pm0.46$&-0.18&-0.20&\\ 
&Kuo10 &-25.53&29.62&12.1&17.6&21.67&27.02&&-0.19&-0.21&\\
& Kuo00 &-25.49&26.34&11.5&14.8&20.20&25.40&&-0.18&-0.19&\\ 
 \hline
& Kuo25&-29.39&27.79&10.2&17.6&17.65&21.80&20.80$\pm 0.30$&-0.17&-0.18&-0.13$\pm $0.06\\
&Kuo20 &-28.26&28.98&10.9&18.1&18.65&23.15&22.27$\pm 0.49$&-0.17&-0.19&\\
$\ ^{72}\rm Ge$&Kuo15 &-27.43&29.68&11.5&18.2&20.08&25.00&23.70$\pm1.80$&-0.18&-0.20\\ 
&Kuo10 &-26.89&29.79&11.9&17.9&21.36&26.64&&-0.19&-0.21&\\
& Kuo00&-26.51&28.39&11.9&16.5&21.79&27.43&&-0.19&-0.21&\\ 
 \hline
&&&&&&$e_\pi$=0.4&0.5&&$e_\nu$=0.4&0.5&\\ 
\hline

& Kuo25&-30.79&26.36&9.7&16.7&29.53&34.75&30.50$\pm 0.30$&-0.21&-0.23&-0.25$\pm $0.06\\
&Kuo20 &-29.41&27.16&10.2&17.0&31.90&37.57&29.00$\pm 2.00$&-0.22&-0.24&\\
$\ ^{74}\rm Ge$&Kuo15 &-28.27&27.92&10.9&17.0&34.36&40.57&30.00$\pm 3.20$&-0.23&-0.25&\\ 
&Kuo10 &-27.49&28.46&11.6&16.9&37.54&44.38&&-0.24&-0.26&\\
& Kuo00&-26.77&28.21&12.0&16.2&41.23&48.91&&-0.25&-0.28&\\ 
\hline
& Kuo25&-31.36&23.63&9.2&14.5&27.63&32.63&27.80$\pm 0.30$&-0.20&-0.22&-0.19$\pm $0.06\\
&Kuo20 &-29.83&24.15&9.6&14.5&29.71&35.10&27.00$\pm2.00$&-0.21&-0.23&\\
$\ ^{76}\rm Ge$&Kuo15 &-28.45&24.77&10.2&14.6&31.98&37.82&26.00$\pm 0.50$&-0.22&-0.24&\\ 
&Kuo10 &-27.36&25.46&11.0&14.5&34.89&41.34&&-0.23&-0.25&\\
& Kuo00 &-26.22&26.10&11.8&14.3&40.22&47.75&&-0.25&-0.27&\\ 
\hline\hline
\end{tabular}

\footnotetext[1]{S. Raman et al., Atom. Nucl. Data Tables $\bf{36}$ (1987) 41.}
\footnotetext[2]{P. Raghavan et al., Atom. Nucl. Data Tables $\bf{42}$ (1989) 189.}
\pagebreak
Table II. The calculated values of Occupation numbers of various sub-shell
orbits for protons and neutrons for $\ ^{70,72,74,76}\rm Ge.$ 

\begin{tabular}{cccccccccc}
\hline\hline
Nucleus&Interaction&\multicolumn{4}{c}{Protons}&\multicolumn{4}{c}{Neutrons} \\
&&$\ 2p_{3/2}$&$\ 2p_{1/2}$&$\ 1f_{5/2}$&$\ 1g_{9/2}$&
$\ 2p_{3/2}$&$\ 2p_{1/2}$&$\ 1f_{5/2}$&$\ 1g_{9/2}$\\ 
 \hline
& Kuo25&0.47&0.34&3.09&0.11&2.01&0.51&4.04&3.44\\
&Kuo20 &0.73&0.41&2.70&0.16&2.22&0.58&3.76&3.44\\
$\ ^{70}\rm Ge$&Kuo15 &1.00&0.47&2.35&0.17&2.46&0.64&3.53&3.37\\ 
&Kuo10 &1.29&0.52&2.03&0.16&2.75&0.72&3.31&3.22\\
& Kuo00 &1.87&0.62&1.42&0.10&3.37&0.96&2.88&2.79\\ 
 \hline
& Kuo25&0.23&0.27&3.50&0.00&2.14&0.53&4.63&4.70\\
&Kuo20 &0.46&0.34&3.11&0.09&2.38&0.61&4.31&4.69\\
$\ ^{72}\rm Ge$&Kuo15 &0.78&0.42&2.67&0.14&2.67&0.70&4.03&4.60\\ 
&Kuo10 &1.10&0.48&2.28&0.14&2.95&0.80&3.79&4.46\\
& Kuo00&1.66&0.57&1.66&0.10&3.42&1.01&3.42&4.15\\ 
 \hline
& Kuo25&0.16&0.22&3.61&0.00&2.40&0.59&5.16&5.85\\
&Kuo20 &0.23&0.27&3.50&0.00&2.61&0.65&4.89&5.86\\
$\ ^{74}\rm Ge$&Kuo15 &0.52&0.36&3.04&0.08&2.90&0.76&4.58&5.75\\ 
&Kuo10 &0.89&0.44&2.55&0.12&3.18&0.90&4.30&5.62\\
& Kuo00&1.54&0.54&1.82&0.10&3.56&1.16&3.90&5.37\\ 
 \hline
& Kuo25&0.11&0.18&3.71&0.00&2.86&0.72&5.54&6.88\\
&Kuo20 &0.16&0.22&3.62&0.00&3.02&0.77&5.38&6.84\\
$\ ^{76}\rm Ge$&Kuo15 &0.28&0.28&3.41&0.03&3.20&0.86&5.16&6.78\\ 
&Kuo10 &0.67&0.37&2.86&0.10&3.42&1.04&4.90&6.64\\
& Kuo00 &1.45&0.50&1.95&0.10&3.72&1.41&4.44&6.44\\ 
\hline\hline

\end{tabular}

\end{document}